\documentclass[conference]{IEEEtran}
\IEEEoverridecommandlockouts

\usepackage{blindtext, graphicx}
\usepackage{cite}
\usepackage[cmex10]{amsmath}
\usepackage{algorithmic}
\usepackage{array}
\usepackage{mdwmath}
\usepackage{mdwtab}
\usepackage{eqparbox}
\usepackage[tight,footnotesize]{subfigure}
\usepackage[font=footnotesize]{subfig}
\usepackage{fixltx2e}
\usepackage{stfloats}
\usepackage{url}
\usepackage{amsthm}
\usepackage{amsfonts}
\usepackage{amsmath}
\usepackage{amssymb}
\usepackage{bbm}
\usepackage{color}

\usepackage{verbatim}

\def\deq{\triangleq}
\def\Integers{\mathbb{Z}}
\def\Reals{\mathbb{R}}
\def\d{\mathrm{d}}
\def\E{\mathbb{E}}
\def\Pr{\mathbb{P}}
\def\cA{\mathcal{A}}
\def\cC{\mathcal{C}}

\def\cN{\mathcal{N}}
\def\cQ{\mathcal{Q}}
\def\cR{\mathcal{R}}
\def\cY{\mathcal{Y}}
\def\sq{\mathsf{q}}
\def\mmse{\mathsf{mmse}}
\def\reg{\mathsf{reg}}
\def\argmin{\operatornamewithlimits{arg\,min}}
\def\1#1{\mathbbm{1}\{#1\}}

\newtheorem{theorem}{Theorem}
\newtheorem{definition}{Definition}
\newtheorem{lemma}{Lemma}
\newtheorem{remark}{Remark}
\newtheorem{corollary}{Corollary}
\newtheorem{proposition}{Proposition}
\newtheorem{assumption}{Assumption}

\begin{document}
\title{On MMSE Estimation from Quantized\\
Observations in the Nonasymptotic Regime}

\author{Jaeho Lee, Maxim Raginsky, and Pierre Moulin
\thanks{The authors are with the Department of Electrical and Computer Engineering and the Coordinated Science Laboratory, University of Illinois, Urbana, IL 61801, USA. E-mails: \{jlee620,maxim,pmoulin\}@illinois.edu.}
\thanks{Research supported in part by DARPA under grant no.\ N66001-13-1-4004 and in part by NSF under CAREER award no.\ CCF-1254041.}}

\maketitle

\begin{abstract} 
	This paper studies MMSE estimation on the basis of quantized noisy observations. It presents nonasymptotic bounds on MMSE regret due to quantization for two settings: (1) estimation of a scalar random variable given a quantized vector of $n$ conditionally independent observations, and (2) estimation of a $p$-dimensional random vector given a quantized vector of $n$ observations (not necessarily independent) when the full MMSE estimator has a subgaussian concentration property. 
	
\end{abstract}



\section{Introduction}
Minimum mean-square error (MMSE) estimation is a fundamental primitive in communications, signal processing, and data analytics. In today's applications, the estimation task is often performed on high-dimensional data collected, and possibly preprocessed, at multiple remote locations. Consequently, attention must be paid to communication constraints and their impact on the MMSE. 

One strategy for reducing the communication burden is to compress the observations using vector quantization (VQ). The idea of quantization for estimation and control can be traced back to the work of Curry \cite{Cur}, who considered in detail the jointly Gaussian case, derived a modification of the Kalman filter for use with quantized inputs, and developed approximation-based schemes for nonlinear systems. Because VQ inevitably introduces loss, it is of interest to characterize the resulting \textit{MMSE regret}, i.e., the difference between the optimal performance achievable with quantized observations and the optimal performance that can be attained without quantization. The problem of designing an optimal vector quantizer to minimize the MMSE regret is equivalent to a noisy source coding problem with the quadratic fidelity criterion, where the compressor acts on the vector of observations and the decompressor generates an estimate of the target random vector. This equivalence was systematically studied by Wolf and Ziv \cite{WolZiv}, who showed that there is no loss of optimality if we first compute the full MMSE estimate and then compress it using an optimal quantizer. Ephraim and Gray \cite{EphGra} later extended this result to a more general class of weighted quadratic distortion functions and gave conditions for convergence of the Lloyd-type iterative algorithm for quantizer design.

The main message of the above works is that the problem of quantizer design for minimum MMSE regret is a \textit{functional compression} problem: the compressed representation of $X$ should retain as much information as possible about the conditional mean $\eta(X) \deq \E[Y|X]$ of the target $Y$ given $X$, also known as the \textit{regression function}. However, there is an interesting tension between the information-theoretic and the \textit{statistical} aspects of the problem: if the MMSE without quantization is sufficiently small, then the dominant contribution to the quantized MMSE should come from the quantization error; in the opposite regime of high-rate quantization, the quantized MMSE should not differ much from its unquantized counterpart.

In this paper, we study both the statistical and the functional-compression aspects of MMSE estimation with quantized observation. In particular, we obtain sharp upper bounds on the MMSE regret in two scenarios:
\begin{itemize}
	\item MMSE estimation of a scalar random variable $Y$, where the vector $X = (X_1,\ldots,X_n)$ of conditionally i.i.d.\ observations is passed through a $k$-ary vector quantizer. In this setting, under mild regularity conditions on the conditional distribution $P_{X_1|Y}$, we obtain nonasymptotic bounds on the MMSE regret. These bounds exhibit two distinct behaviors depending on whether the number of observations $n$ is larger than the square of the codebook size $k^2$. In the asymptotic regime of $n, k \to \infty$, we recover an existing result of Samarov and Has'minskii \cite{SamHas}, who were the first to address this problem.
	\item MMSE estimation of a $p$-dimensional random vector $Y$ when the $n$-dimensional vector of observations $X$ (not necessarily independent) is passed through a $k$-ary VQ. In this setting, we derive an upper bound on the MMSE regret under the assumption that the $\ell_2$ norm of the regression function exhibits subgaussian concentration around its expected value $\E \| \eta(X) \|$. Unlike some of the existing literature on high-resolution quantization for functional compression (e.g., \cite{MisGoyVar}), we do not require smoothness of $\eta(X)$.
\end{itemize}

\noindent{\bf Notation.} We will always denote by $\| \cdot \|$ the $\ell_2$ (Euclidean) norm.  A \textit{$k$-ary quantizer} on the Euclidean space $\mathbb{R}^d$ is a Borel-measurable mapping $q : \mathbb{R}^d \to [k]$, where $[k]$ is shorthand for the set $\{1,\ldots,k\}$. Any such $q$ is characterized by its cells or bins, i.e., the Borel sets $C_j = q^{-1}(\{j\}) = \{ v \in \Reals^d : q(v) = j \}$, $j \in [k]$. The set of all $k$-ary quantizers on $\Reals^d$ will be denoted by $\cQ^d_k$. A \textit{$k$-ary reconstruction function} on $\Reals^d$ is a mapping $f : [k] \to \Reals^d$. Any such $f$ is characterized by its reconstruction points $c_j = f(j) \in \Reals^d$, $j \in [k]$. The set of all $k$-ary reconstruction functions on $\Reals^d$ will be denoted by $\cR^d_k$.
Any finite set of points $\cC = \{c_1,\ldots,c_k\} \subset \Reals^d$ defines a quantizer $q_\cC \in \cQ^d_k$ and a reconstruction function $f_\cC \in \cR^d_k$ by
\begin{align*}
	q_\cC(v) = \argmin_{j \in [k]} \| v - c_j \|, \qquad \forall u \in \Reals^d
\end{align*}
and $f_\cC(j) = c_j$ for all $j \in [k]$. The composite mapping $\sq_\cC \deq f_\cC \circ q_\cC : \Reals^d \to \Reals^d$ is called a \textit{$k$-point nearest-neighbor quantizer} with \textit{codebook} $\cC$. A well-known result is that, for any random vector $V$ with $\E \| V \|^2 < \infty$,
\begin{align*}
	\inf_{f \in \cR^d_k}\inf_{q \in \cQ^d_k} \E \| V - f(q(V)) \|^2 = \inf_{\cC \subset \Reals^d:\, |\cC| = k} \E\|V-\sq_\cC(V)\|^2,
\end{align*}
and the infimum is actually a minimum. We refer the reader to the survey article by Gray and Neuhoff \cite{GraNeu} for more details.

We use the following asymptotic order notation: for two sequences $\{a_m\}$ and $\{b_m\}$, we write $a_m \preceq b_m$ if $a_m = O(b_m)$, and $a_m \asymp b_m$ if $a_m \preceq b_m$ and $b_m \preceq a_m$.

\section{Performance criteria and some basic results}

Suppose two random vectors $X$ in $\Reals^n$ and $Y$ in $\mathbb{R}^p$ are jointly distributed according to a given probability law $P_{XY}$. The MMSE in estimating $Y$ as a function of $X$ is
\begin{align*}
	\mmse(P_{XY}) \deq \inf_f \E \| Y - f(X) \|^2,
\end{align*}	
where the infimum over all Borel-measurable functions $f : \Reals^n \to \Reals^p$ is achieved by the regression function $\eta(x) = \E[Y|X=x]$. We consider the problem of MMSE estimation of $Y$ under the circumstances where only $q(X)$, a quantized version of $X$, is accessible. 
Thus, for each $k \in \Integers^+$, we are interested in the MMSE functional
\begin{align*}
	\mmse_k(P_{XY}) \deq \inf_{f \in \cR^p_k}  \inf_{q \in \cQ^n_k} \E \| Y - f(q(X)) \|^2.
\end{align*}
This problem can be cast as one-shot fixed-rate lossy coding of $Y$ with squared-error distortion when the encoder only has access to $X$ (see, e.g., \cite{WolZiv,EphGra}). We recall some known results that explicitly involve the regression function $\eta(X)$. The first one, due to Wolf and Ziv \cite{WolZiv}, is a useful decomposition of $\mmse_k(P_{XY})$:

\begin{proposition}\label{prop:kMMSE_decomp} For every $k \in \Integers^+$, 
	\begin{align}\label{eq:kMMSE_decomp}
		\mmse_k(P_{XY}) = \mmse(P_{XY}) + \reg_k(P_{XY}),
	\end{align}
	where
	\begin{align*}
		\reg_k(P_{XY}) = \inf_{q \in \cQ^n_k} \E \left\| \E[Y|X] - \E[Y|q(X)] \right\|^2
	\end{align*}
	is the {\em MMSE regret} due to quantization.
\end{proposition}
\begin{remark} {\em A special case of this result for jointly Gaussian $X$ and $Y$ was obtained by Curry \cite[Sec.~2.4]{Cur}.}
\end{remark}

The second result, which was proved by Ephraim and Gray \cite{EphGra} for a more general class of weighted quadratic distortion functions, shows that there is no loss of optimality if we restrict our attention to schemes of the following type: given $X$, we first compute the regression function $\eta(X)$, quantize it using a $k$-ary quantizer, and then estimate $Y$ by its conditional mean given the cell index of $X$.

\begin{proposition}
	\begin{subequations}\label{eq:main}
	\begin{align}
		\reg_k(P_{XY}) &= \inf_{f \in \cR^p_k}\inf_{q \in \cQ^p_k} \E \left\| \eta(X) - f(q(\eta(X)))  \right\|^2 \\
		&= \inf_{\cC \subset \Reals^p:\, |\cC| = k} \E \left\| \eta(X) - \sq_\cC(\eta(X)) \right\|^2.
	\end{align}
\end{subequations}
\end{proposition}

\section{The case of $n$ conditionally independent observations}

We now consider the situation when the coordinates $X_1,\ldots,X_n$ of $X$ are independent and identically distributed (i.i.d.) conditionally on $Y$. To keep things simple, we consider the case when $Y$ is a scalar random variable (i.e., $p=1$), and its marginal distribution $P_Y$ has a probability density function $f_Y$ supported on a compact interval $\cY = [-A,A]$.  In the asymptotic regime as $k \to \infty$ and $n\to\infty$, this setting was investigated by Samarov and Has'minskii \cite{SamHas}, who showed the following under mild regularity conditions:
\begin{itemize}
	\item If $n/k^2 \to \infty$, then the dominant contribution to $\mmse_k(P_{XY})$ comes from the minimum expected distortion
	\begin{align*}
		\inf_{\cC \subset \cY;\, |\cC| = k} \E \left| Y - \sq_\cC(Y) \right|^2 \asymp \frac{1}{k^2} \left(\int_\cY f_Y(y)^{1/3} \d y \right)^3
	\end{align*}
	incurred on $Y$ by any $k$-point quantizer \cite{PanDit,Eli}.
	\item If $n/k^2 \to 0$, then the dominant contribution to $\mmse_k(P_{XY})$ comes from 
	\begin{align*}
		\mmse(P_{XY}) \asymp \frac{1}{n}\int_\cY \frac{f_Y(y)}{I(y)} \d y,
	\end{align*}
	where $I(y)$ is the Fisher information in the conditional distribution $P_{X_1|Y=y}$ \cite{IbrHasPIT} (see below for definitions).
\end{itemize}
The proofs in \cite{SamHas} rely on the deep result, commonly referred to as the \textit{Bernstein--von Mises theorem}, which says that the posterior distribution of the normalized error $\sqrt{nI(Y)}(\eta(X)-Y)$ is asymptotically standard normal (see, e.g., \cite[Sec.~10.2]{Vaa}).

In this section, we establish a finite $n$, finite $k$ version of the results of Samarov--Has'minskii using a recent nonasymptotic generalization of the Bernstein--von Mises theorem due to Spokoiny \cite{Spok_BvM}. Our result requires a number of regularity assumptions. The first two are standard:
\begin{enumerate}
\item[(C.1)] The conditional distributions $P_{X_1|Y=y}$, $y \in \cY$, are dominated by a common $\sigma$-finite measure $\mu$ on the real line. The corresponding log-density
\begin{align*}
	\ell(u,y) \deq \log \frac{\d P_{X_1|Y=y}}{\d \mu}(u)
\end{align*}
is twice differentiable in $y$ for every $u$. We will denote the first and second derivatives with respect to $y$ by $\partial$ and $\partial^2$.
\item[(C.2)] For each $y \in \cY$, the Fisher information \cite{IbrHas}
\begin{align*}
	I(y) \deq -\partial^2 \E\left[\ell(X_1,Y)|Y=y\right]
\end{align*}
exists and is positive.
\end{enumerate}
The remaining assumptions (see \cite[Sec.~5.1]{Spok_AoS}) are slightly stronger than the classical ones. For each $r > 0$, define the local neighborhoods
\begin{align*}
	\cN_r(y) \deq \left\{ y' \in \cY: \sqrt{I(y)}|y'-y| \le r \right\}.
\end{align*}
The regularity assumptions can be split into two groups: identifiability conditions and exponential moment conditions. We start with the former:
\begin{enumerate}
	\item[(I.1)] There are positive constants $r_0 > 0$ and $\delta^* > 0$, such that, for every $r \le r_0$,
	\begin{align*}
		\sup_{y \in \cY}\sup_{y' \in \cN_r(y)} \left|\frac{2D(P_{X_1|Y=y} \| P_{X_1|Y=y'})}{I(y)(y-y')^2}-1\right| \le \delta^* r.
	\end{align*}
	\item[(I.2)] There is a constant $b > 0$, such that, for every $r > 0$,
	\begin{align*}
		\inf_{y \in \cY}\inf_{y' \in \partial \cN_r(y)} \frac{D(P_{X_1|Y=y}\|P_{X_1|Y=y'x})}{I(y) (y-y')^2} \ge b.
	\end{align*}
\end{enumerate}
The exponential moment assumptions pertain to the random variables
\begin{align*}
	\zeta_y(X_1,y') \deq \ell(X_1,y') - \E[\ell(X_1,y')|Y=y], \quad y,y' \in \cY
\end{align*}
[note that $\zeta_y(Z_1,y) \equiv \ell(X_1,y)$]. Let
\begin{align*}
	\xi_y(X_1,y') \deq \frac{\partial}{\partial w} \zeta_y(X_1,w)\Big|_{w=y'}.
\end{align*}
\begin{enumerate}
\item[(E.1)] There exist constants $g_1 > 0$ and $v_0 > 0$, such that
\begin{align*}
	\sup_{y \in \cY}\log \E \Bigg[\exp\left(\frac{\lambda \xi_y(X_1,y)}{\sqrt{I(y)}}\right)\Bigg|Y=y\Bigg] \le \frac{v^2_0 \lambda^2}{2}
\end{align*}
for all $|\lambda| \le g_1$.
\item[(E.2)] There exists a constant $\omega^* > 0$, such that, for every $r \le r_0$,
\begin{align*}
&\!\!\!\!\!\!\!\!\!\!\!\!\!\!\!\!\sup_{y \in \cY}\sup_{y' \in \cN_r(y)}\log \E\Bigg[ \exp\left(\frac{\lambda [\xi_y(X_1,y')-\xi_y(X_1,y)]}{\omega^* r \sqrt{I(y)}}\right)\Bigg|Y=y\Bigg] \\
	&\qquad \qquad \le \frac{v^2_0 \lambda^2}{2}
\end{align*}
for all $|\lambda| \le g_1$.
\item[(E.3)] For every $r > 0$ there exists $g_1(r) > 0$, such that
\begin{align*}
\!\!\!\!\!	\sup_{y \in \cY}\sup_{y' \in \cN_r(y)}\log \E\Bigg[\exp\left(\frac{\lambda \xi_y(X_1,y)}{\sqrt{I(y)}}\right)\Bigg|Y=y\Bigg] \le \frac{v^2_0 \lambda^2}{2}.
\end{align*}
for all $|\lambda| \le g_1(r)$.
\end{enumerate}
For example, in the case of additive Gaussian noise, i.e., when $P_{X_1|Y=y} = N(y,\sigma^2)$ for all $y \in \cY$,  it is easy to verify that all of these assumptions are met. 

We are now ready to state and prove the main result of this section. Since both $Y$ and $\eta(X)$ are supported on the bounded interval $[-A,A]$, there is no loss of generality in restricting our attention only to nearest-neighbor quantizers $\sq_\cC$ with codebooks $\cC = \{y_1,\ldots,y_k\} \subset [-A,A]$. Moreover, we can assume that $\cC$ is ordered in such a way that $-A \equiv y_0 \le y_1 < y_2 < \ldots < y_k \le y_{k+1} \equiv A$. Given such an ordered $\cC$, we define
\begin{align*}
	\Delta_\cC \deq \max_{0 \le j \le k} (y_{j+1}-y_{j}).
\end{align*}

\begin{theorem}\label{thm:quantizer_BvM} Suppose that Assumptions (C.1)--(C.2), (I.1)--(I.2), and (E.1)--(E.3) hold. Suppose also that $\log f_Y$ is Lipschitz on $[-A,A]$. Then there exists a constant $L > 0$ that depends only on the constants in the above assumptions, such that, for any $k$-point nearest-neighbor quantizer $\sq_\cC$ with $\cC \subset \cY$, we have
	\begin{align}
	&	\left| \E\left|\eta(X)-\sq_\cC(\eta(X))\right|^2 - \E\left|Y-\sq_\cC(Y)\right|^2\right|  \nonumber\\
	& \le L\Delta^2_\cC \min \Bigg\{1, \frac{1}{\Delta_\cC\sqrt{n}} \Bigg(\E\left[\frac{1}{\sqrt{I(Y)}}\right]+\sqrt{\mmse(P_{XY})}\Bigg)\Bigg\}. \label{eq:quantizer_BvM}
	\end{align}
\end{theorem}
\noindent In the additive Gaussian noise case, the bound \eqref{eq:quantizer_BvM} becomes
\begin{align*}
&\left| \E\left|\eta(X)-\sq_\cC(\eta(X))\right|^2 - \E\left|Y-\sq_\cC(Y)\right|^2\right| \\
& \qquad \qquad  \le L\Delta^2_\cC \min \Bigg\{1, \frac{\sigma}{\Delta_\cC\sqrt{n}}\Bigg\},
\end{align*}
where $\sigma^2$ is the noise variance.

\begin{IEEEproof} The idea of the proof of our nonasymptotic result is actually rather simple, unlike that of Samarov and Has'minskii \cite{SamHas}, which requires a number of delicate asymptotic approximations and several fairly tedious integrations. For any collection $\cC = \{y_1,\ldots,y_k\}$ of $k$ reconstruction points, define the function $e_\cC : \cY \to \Reals^+$ by
	\begin{align*}
		e_\cC(y) \deq \min_{j \in [k]} (y-y_j)^2.
	\end{align*}
Then a simple calculation shows that
\begin{align}\label{eq:error_smoothness}
	\left|e_\cC(y) - e_\cC(y')\right| \le \min \left\{ 2\Delta^2_\cC, 2\Delta_\cC |y-y'|\right\},
\end{align}
for all $y,y' \in \cY$. The expected reconstruction error of the nearest-neighbor quantizer $\sq_\cC$ can be written as
	\begin{align*}
		\E \left|\eta(X) - \sq_\cC(\eta(X)) \right|^2 &= \E[e_\cC(\eta(X))].
	\end{align*}
Using the law of iterated expectation and the smoothness estimate \eqref{eq:error_smoothness}, we obtain
\begin{align}
	&\left| \E[e_\cC(\eta(X))] - \E[e_\cC(Y)] \right|\nonumber\\
	 &\qquad \le \E\left| \E[e_\cC(\eta(X))-e_\cC(Y)|Y]\right| \nonumber\\
	&\qquad \le 2\Delta^2_\cC \E \min \left\{ 1, \frac{1}{\Delta_\cC} \E\big[|\eta(X)-Y| \big|Y\big] \right\}. \label{eq:BvM_bound_1}
\end{align}
We now invoke Spokoiny's nonasymptotic Bernstein--von Mises theorem \cite{Spok_BvM}. Let $Z_n \deq \eta(X) - (Y + G_n(X,Y))$, where
\begin{align*}
	G_n(X,Y) \deq \frac{1}{n I(Y)} \sum^n_{i=1} \partial \ell(X_i,Y), 
\end{align*}
and for any $L > 0$ consider the event
\begin{align*}
	\cA^L_n(Y) \deq \left\{ \sqrt{nI(Y)}|Z_n| \le L\left(\frac{\log n}{n}\right)^{1/4}\right\},
\end{align*}
Then there exists a choice $L = L_0$ that depends only on the constants in the regularity conditions, such that
\begin{align}\label{eq:BvM}
	\Pr\big[\cA_n(Y)\big|Y\big] \ge 1-\frac{C}{n},
\end{align}
where $\cA_n(Y) \equiv \cA^{L_0}_n(Y)$, and $C > 0$ is an absolute constant \cite[Sec.~2.4.3]{Spok_BvM}. Therefore,
\begin{align}
	&\E\big[|\eta(X)-Y| \big|Y\big] \nonumber\\
	&\stackrel{{\rm (a)}}{\le} \E\big[|Z_n| \big|Y\big] + \E\big[|G_n(X,Y)|\big|Y\big] \nonumber\\
	&= \E\big[\1{\cA_n(Y)}|Z_n| \big|Y\big] + \E\big[\1{\cA^c_n(Y)}|Z_n| \big|Y\big] \nonumber\\
	&\qquad \qquad + \E\big[|G_n(X,Y)|\big|Y\big] \nonumber\\
	&\stackrel{{\rm (b)}}{\le} \frac{L_0}{\sqrt{nI(Y)}}+ \sqrt{\frac{C}{n}\E\big[|Z_n|^2\big|Y\big]}  + \E\big[|G_n(X,Y)|\big|Y\big] \nonumber\\
	&\stackrel{{\rm (c)}}{\le} \frac{L_0}{\sqrt{nI(Y)}} + \sqrt{\frac{2C}{n}\E[|Y-\eta(X)|^2\big|Y\big]} \nonumber\\
	& \,\,  + \sqrt{\frac{2C}{n} \E\big[\big|G_n(X,Y)|^2\big|Y\big]} + \E\big[|G_n(X,Y)|\big|Y\big], \label{eq:BvM_bound}
\end{align}
where (a) follows from the triangle inequality, (b) from \eqref{eq:BvM} and Cauchy--Schwarz, and (c) again from the triangle inequality. Now, since $\E[\partial \ell(X_1,Y)|Y] = 0$ and ${\rm Var}[\partial \ell(X_1,Y)|Y] = I(Y)$ \cite{IbrHas}, we have
\begin{align*}
	\E\big[|G_n(X,Y)|^2\big|Y\big] &= {\rm Var}[G_n(X,Y)|Y] = \frac{1}{nI(Y)} 
\end{align*}
and
\begin{align*}
	\E\big[|G_n(X,Y)|\big|Y\big] &\le \frac{1}{\sqrt{nI(Y)}}.
\end{align*}
Substituting these estimates into \eqref{eq:BvM_bound} gives
\begin{align*}
	\E\big[|\eta(X)-Y| \big|Y\big] &\le \frac{L_0+1}{\sqrt{nI(Y)}}  + \frac{\sqrt{2C}}{n\sqrt{I(Y)}} \\
	& \qquad \qquad + \sqrt{\frac{2C\, \E\big[|Y-\eta(X)|^2\big|Y\big]}{n}}.
\end{align*}
Plugging this bound into \eqref{eq:BvM_bound_1}, using Jensen's inequality, and simplifying, we get \eqref{eq:quantizer_BvM}.
\end{IEEEproof}

\begin{corollary} Under the same assumptions as in Theorem~\ref{thm:quantizer_BvM},
	\begin{align}
	&	\left|\reg_k(P_{XY})-\inf_{\cC \subset \cY;\, |\cC| \le k} \E|Y-\sq_\cC(Y)|^2\right| \nonumber\\
	& \quad \preceq \min\left\{ \frac{1}{k^2}, \frac{1}{k\sqrt{n}}\left(\E\left[\frac{1}{\sqrt{I(Y)}}\right] + \sqrt{\mmse(P_{XY})}\right)\right\}. \label{eq:quantizer_BvM_2}
	\end{align}
\end{corollary}
\begin{remark} {\em Using the information inequality \cite{IbrHasPIT}
\begin{align*}
	\mmse(P_{XY}) \ge \frac{1}{n}\E\left[\frac{1}{I(Y)}\right]
\end{align*}
and Jensen's inequality, the bound \eqref{eq:quantizer_BvM_2} can be weakened to
\begin{align*}
&	\left|\reg_k(P_{XY})-\inf_{\cC \subset \cY;\, |\cC| \le k} \E|Y-\sq_\cC(Y)|^2\right| \nonumber\\
& \qquad\qquad \preceq \min\left\{ \frac{1}{k^2}, \frac{\sqrt{\mmse(P_{XY})}}{k}\right\}.
\end{align*}
}
\end{remark}
\begin{proof} Let $\cC^* = \{y^*_1,\ldots,y^*_k\}$ be the reconstruction points of an optimal $k$-point quantizer for $Y$ arranged in increasing order. From the work of Panter and Dite \cite{PanDit}, we know that these points should be chosen in such a way that
	\begin{align*}
		\int^{y^*_{j+1}}_{y^*_j} f_Y(y)^{1/3} \d y = \frac{c_j}{k}\int_\cY f_Y(y)^{1/3}\d y, 
	\end{align*}
	for all $j = 0, 1,\ldots,k-1$, where $c_j = 1/2$ for $j=0$ and $1$ otherwise. Therefore, $\Delta_{\cC^*} \preceq 1/k$. Using this and the definition of the MMSE regret in \eqref{eq:quantizer_BvM}, we get \eqref{eq:quantizer_BvM_2}.
\end{proof}
Observe that the value of the right-hand side of \eqref{eq:quantizer_BvM_2} is determined by whether the number of observations $n$ is larger or smaller than $k^2$. This agrees with the asymptotic results of Samarov and Has'minskii \cite{SamHas}.

\section{A high-resolution bound for functional compression}

We now consider a general setting of $p$-dimensional $Y$ and $n$-dimensional $X$ without any conditional independence assumptions. Instead, we focus on the scaling of the MMSE regret $\reg_k(P_{XY})$ with $k$, while the values of $n$ and $p$ stay fixed. Proposition~\ref{prop:kMMSE_decomp} shows that the MMSE regret is precisely the minimum expected distortion attainable by $k$-ary quantization of $X$ in the problem of \textit{functional compression} of the regression function $\eta(X)$. The results we present in this section require only some regularity assumptions on the $\ell_2$ norm $\| \eta(X) \|$:

\begin{assumption}\label{as:4th} The random variable $\| \eta(X) \|$ has a finite fourth moment: $\E \| \eta(X) \|^4 < \infty$.
\end{assumption}
\begin{assumption}\label{as:SG} The random variable $\| \eta(X) \|$ is subgaussian: there exists a positive constant $v > 0$, such that
	\begin{align*}
		\log\E\left[e^{\lambda (\| \eta(X) \| - \E \|\eta(X) \|)}\right] \le \frac{v\lambda^2}{2}, \qquad \forall \lambda \in \Reals.
	\end{align*}
\end{assumption}
One sufficient condition for Assumption~\ref{as:4th} to hold is for $\| Y \|$ to have a finite fourth moment. Indeed, in that case, using Jensen's inequality, we have $\E \| \eta(X) \|^4 = \E \| \E[Y|X] \|^4 \le \E \| Y \|^4 < \infty$. As for Assumption~\ref{as:SG}, it will be met, for example, if the regression function $\eta$ is Lipschitz, i.e., if there exists some finite constant $L > 0$, such that
$$
\left\|\eta(x) - \eta(x')\right\| \le L\|x - x'\|, \qquad \forall x,x' \in \Reals^n
$$
and if $X$ is a Gaussian random vector with a nonsingular covariance matrix \cite{RagSas}.

\begin{theorem} Suppose Assumption~\ref{as:4th} holds. Then
	\begin{align}\label{eq:hi_res_achievability_1}
		\reg_k(P_{XY}) \preceq \left(\E\| \eta(X) \|^2 \E\|\eta(X)\|^4\right)^{2/3}k^{-2/3p}.
	\end{align}
If Assumption~\ref{as:SG} also holds, then
	\begin{align}\label{eq:hi_res_achievability_2}
	&	\reg_k(P_{XY}) \nonumber\\
	& \preceq \inf_{r > \E\|\eta(X)\|} \Big\{ r^2 k^{-2/p}  + \sqrt{\E \| \eta(X) \|^4} e^{-(r-\E\|\eta(X)\|)^2/4v} \Big\}.
	\end{align}
\end{theorem}
\begin{remark}{\em Here, $p$ can be replaced by a suitable intrinsic dimension of the support of $\eta(X)$ (e.g., the rate-distortion dimension \cite{KawDem}). For example, if $\eta(X)$ is linear, i.e., $\E[Y|X]=AX$ for some deterministic matrix $A \in \Reals^{p \times n}$, then we can replace $p$ by ${\rm rank}(A)$. Moreover, using a suboptimal value of $r$, we can weaken the bound in  \eqref{eq:hi_res_achievability_2} to
	\begin{align*}
		\reg_k(P_{XY}) \preceq \frac{\log k}{k^{2/p}},
	\end{align*}
where the hidden constant depends on $p$, on the first and fourth moments of $\| \eta(X) \|$, and on the subgaussian constant $v$. Apart from the logarithmic factor, this scaling of the MMSE regret agrees with the high-resolution approximation for VQ \cite{GraNeu} and with the Shannon lower bound \cite{LinZam,KawDem}.}
\end{remark}
\begin{IEEEproof} From Assumption~\ref{as:4th} and from Jensen's inequality, it follows that the first and second moments of $\| \eta(X) \|$ are also finite.

Fix a positive real constant $r > 0$, which will be optimized later. A simple volumetric estimate shows that the $\ell^2$ ball of radius $r$ in $\Reals^p$ can be covered by at most $\left(1+\frac{2r}{\epsilon}\right)^p$ balls of radius $\epsilon$.  So, for a given $k \in \Integers^+$, we can cover the radius-$r$ ball by $k$ balls of radius $\epsilon \asymp r k^{-1/p}$. Let $\{y_i\}_{i=1}^k$ be the centers of these $k$ balls. We now construct a quantizer $q^{(r)} \in \cQ^p_{k+1}$ as follows:
\begin{align*}
	q^{(r)}(x) &= \begin{cases}
	\displaystyle\argmin_{j \in [k]} \| \eta(x) - y_j \|^2, & \text{if } \| \eta(x) \| \leq r \\
	k+1, &\text{otherwise}.
\end{cases}
\end{align*}
For this quantizer, we have
\begin{align*}
	\E \| \eta(X) - \E[\eta(X)|q^{(r)}(X)] \|^2 = T_1 + T_2,
\end{align*}
where
\begin{align*}
T_1 &\triangleq \mathbb{E}\left[\1{\|\eta(X)\| \leq r} \|\eta(X) - \E[\eta(X)|q^{(r)}(X)]\|^2\right]\\
T_2 &\triangleq \mathbb{E}\left[\1{\|\eta(X)\| > r} \|\eta(X) - \E[\eta(X)|q^{(r)}(X)]\|^2\right].
\end{align*}
For the first term, we have
\begin{align*}
T_1 \leq 4\epsilon^2\Pr\left[\|\eta(X)\| \leq r\right] \leq 4\epsilon^2 \asymp r^2 k^{-2/p},
\end{align*}
where the first inequality is true since $\eta(X)$ and $\E[\eta(X)|q^{(r)}(X)]$ are inside the same $\epsilon$-ball in the covering for all $X$ such that $q^{(r)}(X) \in [k]$.
For the second term,
\begin{align*}
T_2 &\leq \sqrt{\Pr\left[\|\eta(X)\| > r\right]\mathbb{E}\|\eta(X) - \E[\eta(X)|q^{(r)}(X)\|^4}\\
&\leq \sqrt{\Pr[\|\eta(X)\| > r]}\sqrt{\mathbb{E}(\|\eta(X)\| + \|\E[\eta(X)|q^{(r)}(X)]\|)^4}\\
&\leq \sqrt{\Pr[\|\eta(X)\| > r]}\sqrt{8\,\mathbb{E}\|\eta(X)\|^4 + 8\,\mathbb{E}\|\E[\eta(X)|q^{(r)}]\|^4}\\
&\leq \sqrt{\Pr[\|\eta(X)\| > r]}\sqrt{16\,\mathbb{E}\|\eta(X)\|^4},
\end{align*}
where the first line is by Cauchy--Schwarz inequality, and the remaining steps follow from monotonicity and convexity. 

By Markov's inequality,
\begin{align*}
	\Pr[\|\eta(X)\| > r] = \Pr[\|\eta(X)\|^2 > r^2] \le \frac{\E \|\eta(X)\|^2}{r^2}.
\end{align*}
Therefore,
\begin{align*}
	 \reg_{k+1}(P_{XY}) &\le	\E \| \eta(X) - \E[\eta(X)|q^{(r)}] \|^2\nonumber\\
	& \asymp r^2 k^{-2/p} + \frac{\sqrt{\E \| \eta(X)\|^2 \E \| \eta(X) \|^4}}{r}.
\end{align*}
Optimizing over all $r > 0$, we get \eqref{eq:hi_res_achievability_1}. Now suppose Assumption~\ref{as:SG} also holds. Let $r = \E \| \eta(X) \| + t$ for some $t > 0$. Since $\| \eta(X) \|$ is subgaussian, the Chernoff bounding technique gives
\begin{align*}
	\Pr\left[\| \eta(X) \| > r\right] &= \Pr\left[\|\eta(X)\| - \E\|\eta(X)\| > t\right] \\
	&\le e^{-t^2/2v} \\
	&= e^{-(r-\E\|\eta(X)\|)^2/2v}
\end{align*}
(see, e.g., \cite[Chap.~3]{RagSas}). Thus, we have
\begin{align*}
 \reg_{k+1}(P_{XY}) &\le	\E \| \eta(X) - \E[\eta(X)|q^{(r)}] \|^2\nonumber\\
&  \asymp r^2k^{-2/p} + \sqrt{\E \| \eta(X) \|^4} e^{-(r - \E \| \eta(X) \|)^2/4v}.
\end{align*}
for all $r > \E \| \eta(X) \|$. Optimizing over $r$, we get \eqref{eq:hi_res_achievability_2}.
\end{IEEEproof}

\section*{Acknowledgments} The authors would like to thank Tam\'as Linder and Vladimir Spokoiny for helpful discussions.

\ifCLASSOPTIONcaptionsoff
  \newpage
\fi


\end{document}